\documentclass[showpacs,aps,graphicx,twocolumn]{revtex4}
\usepackage{graphicx}

\begin{document}

\title{Efficient W state entanglement concentration using quantum-dot and  optical microcavities}

\author{Yu-Bo Sheng,$^{1,2}$\footnote{Email address:
shengyb@njupt.edu.cn} Lan Zhou,$^{1,2}$, Chuan Wang,$^{3}$, and
Sheng-Mei Zhao,$^{1,2}$ }
\address{$^1$ Institute of Signal Processing  Transmission, Nanjing
University of Posts and Telecommunications, Nanjing, 210003,  China\\
$^2$Key Lab of Broadband Wireless Communication and Sensor Network
 Technology,
 Nanjing University of Posts and Telecommunications, Ministry of
 Education, Nanjing, 210003,
 China\\
 $^3$School of Science, Beijing University of Posts and
Telecommunications, Beijing, 100876, China\\}
\date{\today}

\begin{abstract}
We present an entanglement concentration protocols (ECPs) for
less-entangled W state with quantum-dot and microcavity coupled
system. The present protocol uses the quantum nondemolition
measurement on the spin parity to construct the parity check gate.
Different from other ECPs, this less-entangled W state with
quantum-dot and microcavity coupled system can be concentrated with
the help of some single photons. The whole protocol can be repeated to get a higher success probability.
It may be useful in current quantum information processing.
\end{abstract}

 \maketitle

\section{introduction}
Entanglement plays an important role in  quantum information
processing \cite{book,rmp}. It has many practical applications, such
as quantum teleportation \cite{teleportation,cteleportation}, quantum
cryptograph \cite{Ekert91,cryp}, quantum dense
coding \cite{densecoding}, quantum secure direct
communication \cite{QSDC1,QSDC2,QSDC3} and other quantum information
protocols \cite{QSTS1,QSTS2,QSTS3,QSS1,QSS2,QSS3}.  In order to
achieve such tasks, the legitimate uses, say the sender Alice and
the receiver Bob should first share the entanglement in distant
locations. In a long-distance quantum communication, quantum
repeaters are unusually used to sent the quantum signals over an
optical fiber or a free space \cite{repeater1,DLCZ}. Unfortunately,
in a practical transmission, the entangled quantum system cannot
avoid the channel noise from the environment. It will make the
quantum system decoherence. That is, a maximally entangled state will
become a less-entangled state or a mixed entangled state.

Quantum concentration is to distill some maximally entangled state
from an ensemble in a pure less-entangled
state \cite{C.H.Bennett2,swapping1,swapping2,zhao1,zhao2,Yamamoto1,Yamamoto2,
bose,cao1,cao2,shengpra2,shengpra3,shengqic,wangxb,shengpla,zhang,dengconcentration,shengwstateconcentration,shengsinglephotonconcentration}.
In 1996, Bennett \emph{et al.} proposed an entanglement
concentration protocol (ECP), which is called Schmidit  projection
method \cite{C.H.Bennett2}. Bose \emph{et al.} proposed an ECP based
on the swapping \cite{bose}. ECPs based on linear optical elements
were proposed by Yamamoto \emph{et al.} and Zhao \emph{et al.},
respectively \cite{zhao1,zhao2,Yamamoto1,Yamamoto2}. In 2008, an ECP
based on cross-Kerr nonlinearity was proposed \cite{shengpra2}. In
2011, an ECP based on the quantum-dot in an optical cavity was
proposed. This kind of ECPs are all used to concentrate a
less-entangled state
$\alpha|0\rangle|0\rangle+\beta|1\rangle|1\rangle$ to a maximally
entangled state
$\frac{1}{\sqrt{2}}(|0\rangle|0\rangle+|1\rangle|1\rangle)$. Unusually, in each concentration step, they choose two similar
copies of less-entangled states and after performing the ECP, at least one pair of maximally entangled state can be obtained
with certain probability. The most advantage of this ECP is that
they do not need to know the exact coefficients $\alpha$ and
$\beta$. On the other hand, there is another kind of ECP which can
be used to concentration a less-entangled W state into a maximally
entangled W state. For example, in 2003, Cao  and Yang proposed an
ECP for W-class state using joint unitary transformation\cite{cao}.
Zhang \emph{et al.} proposed an ECP with the help of collective
Bell-state measurement \cite{zhanglihua}. The ECPs for a special
less-entangled W state were proposed in both linear optical system
and cavity QED system \cite{wanghf1,wanghf2}. In 2011, Yildiz
proposed an optimal ECP for asymmetric W states of the
form \cite{yildiz}
\begin{eqnarray}
\frac{1}{\sqrt{2}}|001\rangle+\frac{1}{2}|010\rangle+\frac{1}{2}|100\rangle,\nonumber\\
\frac{1}{2}|001\rangle+\frac{1}{2}|010\rangle+\frac{1}{\sqrt{2}}|100\rangle.
\end{eqnarray}

On the other hand, the ECPs described above are unusually based on
the universal qubit say $|0\rangle$ and $|1\rangle$ \cite{C.H.Bennett2} or the optical
system, which $|0\rangle\equiv|H\rangle$ and $|1\rangle=|V\rangle$ \cite{Yamamoto1,Yamamoto2,zhao1,zhao2}.
Here $|H\rangle$ and $|V\rangle$ represent the horizontal and
vertical polarization of the photon, respectively. Recently, there
is a novel candidate  for qubit which is a single spin coupled to an
optical microcavity based on a charged
qantum-dot \cite{cavity,hu1,hu2,hu3,hu4}. For example, Hu \emph{et
al.} proposed an deterministic photon entangler using a charged
quantum dot inside a microcavity\cite{hu1}. They   also
proposed an entanglement beam splitter and discussed the
loss-resistant state teleportation and entanglement swapping using a
quantum dot spin in an optical microcavity \cite{hu2,hu3,hu4}. Bonato
\emph{et al.} discussed the CNOT gate and Bell-state analysis in the
weak-coupling cavity QED regime \cite{cavity}.  Wang
\emph{et al.} proposed  two entanglement purification protocols
based on the hybrid entangled state using quantum-dot and
microcavity coupled system \cite{wangc2,wangc3}. Recently, a efficient quantum repeater protocol
was proposed \cite{wangtiejun}. Inspired by the
novel works of Hu and Wang, we propose an ECP for the less-entangled W
state exploiting the quantum-dot and microcavity coupled system. In
this protocol, the less-entangled W state of the spin in the cavity
QED system can be concentrated into a maximally entangled W state
with some ancillary single photons. This protocol is quite different
from the others. First, we can concentrate the arbitrary
less-entangled W state. Second, we do not need two copies of
less-entangled pairs. Third, the ECP can be performed between
different degrees of freedoms, that is we use the single photons
 to concentrate the less-entangled state in spin.  Fourth,
by repeating this ECP, it can reach a higher success probability.

This paper is organized as follows: In Sec. II, we describe the
theoretical model for our ECP. We call it a hybrid parity check gate based on
photon and electron coupled systems. In Sec. III, we explain our ECP
based on the parity check gate. In Sec. IV, we discuss the
efficiency and errors on a practical implementation. In Sec. V, we
present a discussion and summary.

\section{hybrid parity check gate}
Before we start to explain this ECP. We first describe the basic
element for this protocol. It is also shown in Refs. \cite{hu2,cavity,wangc2,wangc3}. As shown in Fig. 1, the system is
composed of a single charged quantum-dot in micropillar
microcavites. The charge exciton consists of two electrons bound in
one hole and the excitation with negative charges can created by the
optical excitation of the system. Therefore, if we consider a photon
entrances into the cavity from the input mode and it will interact with the electron in
the coupling cavity. Interestingly, the left circularly polarized
photon $|L\rangle$ only couples with the electron in the spin up
state $|\uparrow\rangle$ to the exciton $X^{-}$ in the state
$|\uparrow\downarrow\Uparrow\rangle$ because of the  Pauli's
exclusion principle for two electrons. On the other hand, the right
circularly polarized photon $|R\rangle$ only couples with the
electron of the spin down $|\downarrow\rangle$ in the state
$|\downarrow\uparrow\Downarrow\rangle$. Here the $|\Uparrow\rangle$
and $|\Downarrow\rangle$ are the spin direction of the heavy hole
spin state. In Ref. \cite{hu2}, Hu \emph{ et al.} discussed that such system
essentially is an entanglement beam splitter which directly splits
an initial hybrid product state of photon and spin into two
entangled states via transmission and refection in a deterministic
way. They denoted the transmission and reflection operators as

\begin{eqnarray}
\hat{t}&=&|R\rangle\langle R|\otimes
|\uparrow\rangle\langle\uparrow|+|L\rangle\langle L|\otimes
|\downarrow\rangle\langle\downarrow|,\nonumber\\
\hat{r}&=&|R\rangle\langle R|\otimes
|\uparrow\rangle\langle\uparrow|+|L\rangle\langle L|\otimes
|\downarrow\rangle\langle\downarrow|.
\end{eqnarray}
From Fig. 1, we consider a photon is in the state
$|R^{\uparrow}\rangle$ with $s_{z}=+1$ and the electron spin is
$|\uparrow\rangle$. Here the superscript $|\uparrow\rangle$ means
the photon's propagation direction is along the z axis. Both the
polarization of the photon and propagation direction are flipped
into $|L^{\downarrow}\rangle$. In the same way, the photon and
electron interaction in quantum dot and microcavity coupled systems
can be fully described as
\begin{eqnarray}
&&|R^{\uparrow},\uparrow\rangle\rightarrow|L^{\downarrow},\uparrow\rangle,
\hspace{8mm}|R^{\downarrow},\uparrow\rangle\rightarrow
-|R^{\downarrow},\uparrow\rangle,\nonumber\\
&&|R^{\uparrow},\downarrow\rangle\rightarrow
-|R^{\uparrow},\downarrow\rangle,
\hspace{5mm}|R^{\downarrow},\downarrow\rangle\rightarrow
|L^{\uparrow},\downarrow\rangle,\nonumber\\
&&|L^{\uparrow},\uparrow\rangle\rightarrow
-|L^{\uparrow},\uparrow\rangle,\hspace{5.5mm}|L^{\downarrow},
\uparrow\rangle\rightarrow|R^{\uparrow},\uparrow\rangle,\nonumber\\
&&|L^{\uparrow},\downarrow\rangle\rightarrow|R^{\downarrow},\downarrow\rangle,\hspace{8mm}
|L^{\downarrow},\downarrow\rangle\rightarrow
-|L^{\downarrow},\downarrow\rangle.
\end{eqnarray}
\begin{figure}[!h]
\begin{center}
\includegraphics[width=3.5cm,angle=0]{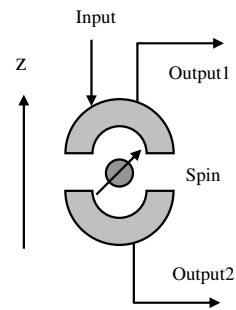}
\caption{A schematic drawing of the hybrid parity check gate for our ECP. The
quantum-dot spin is coupled in the  optical microcavity. The input and output represent the input and output ports of a photon.
This setup can split a photon-spin product state into two constituent hybrid  photon-spin entangled state. One is
in the output1 mode and another is in the output2 mode. }
\end{center}
\end{figure}
So if the initial input state is the photon-spin product state
$\frac{1}{\sqrt{2}}(|R\rangle+|L\rangle)\otimes\frac{1}{\sqrt{2}}(|\uparrow\rangle+|\downarrow\rangle)$,
it will be converted into the two constituent hybrid entangled state
$\frac{1}{\sqrt{2}}(|R\rangle|\uparrow\rangle+|L\rangle|\downarrow\rangle)$
in the transmission port, say output2 mode  and
$\frac{1}{\sqrt{2}}(|R\rangle|\downarrow\rangle+|L\rangle|\uparrow\rangle)$
in the reflection port, say output1 mode respectively, with the success probability
of 100\%, in principle. In the following, we denote the transmission port as output2 mode and
the reflection port as output1 for simple, shown in Fig. 1. A parity check gate has been widely in
current quantum processing. Entanglement purification and concentration all need such elements.
An optical parity check gate, such as polarization beam splitter, can convert the product state
$\frac{1}{\sqrt{2}}(|H\rangle+|V\rangle)\otimes\frac{1}{\sqrt{2}}(|H\rangle+|V\rangle)$ into two constituent
entangled state $\frac{1}{\sqrt{2}}(|H\rangle|H\rangle+|V\rangle|V\rangle)$ and  $\frac{1}{\sqrt{2}}(|H\rangle|V\rangle+|V\rangle|H\rangle)$.
Compared with the polarization beam splitter
in optical system, it  essentially acts the same role of the parity
check gate, but with different degrees of freedom.  So we call it a
hybrid parity check gate.

\section{ECP for less-entangled W state}
Now we start to explain our protocol. From Fig. 2, the
less-entangled W state are shared by Alice, Bob and Charlie. It can
be written as:
\begin{eqnarray}
|\Phi\rangle_{123}&=&\alpha_{1}|\downarrow\rangle_{1}|\uparrow\rangle_{2}|\uparrow\rangle_{3}+\alpha_{2}|\uparrow\rangle_{1}|\downarrow\rangle_{2}|\uparrow\rangle_{3}\nonumber\\
&+&\alpha_{3}|\uparrow\rangle_{1}|\uparrow\rangle_{2}|\downarrow\rangle_{3},\label{W
state}
\end{eqnarray}

Here $|\alpha_{1}|^{2}+|\alpha_{2}|^{2}+|\alpha_{3}|^{2}=1$ and subscripts 1, 2 and 3 mean spin 1, spin 2 and spin 3
respectively.  Suppose that the three parties know the initial
coefficients $\alpha_{1}$, $\alpha_{2}$, and $\alpha_{3}$.  Alice first prepare a
single photon of the form
\begin{eqnarray}
|\Phi\rangle_{P1}&=&\frac{\alpha_{1}}{\sqrt{\alpha_{1}^{2}+\alpha_{2}^{2}}}|R\rangle_{1}+\frac{\alpha_{2}}{\sqrt{\alpha_{1}^{2}+\alpha_{2}^{2}}}|L\rangle_{1}.\label{auxiliary1}
\end{eqnarray}
The subscript $P1$ means the photon coupled with the  spin 1. She
sends the state $|\Phi\rangle_{P1}$ to the cavity from the input mode. The initial
less-entangled W state combined with the single photon evolve as
\begin{eqnarray}
|\Psi\rangle&=&|\Phi\rangle_{123}|\Phi\rangle_{P1}
=(\alpha_{1}|\downarrow\rangle_{1}|\uparrow\rangle_{2}|\uparrow\rangle_{3}+\alpha_{2}|\uparrow\rangle_{1}|\downarrow\rangle_{2}|\uparrow\rangle_{3}\nonumber\\
&+&\alpha_{3}|\uparrow\rangle_{1}|\uparrow\rangle_{2}|\downarrow\rangle_{3})\nonumber\\
&(&\frac{\alpha_{1}}{\sqrt{\alpha_{1}^{2}+\alpha_{2}^{2}}}|R^{\downarrow}\rangle_{1}+\frac{\alpha_{2}}{\sqrt{\alpha_{1}^{2}+\alpha_{2}^{2}}}|L^{\downarrow}\rangle_{1})\nonumber\\
&=&\frac{\alpha_{1}^{2}}{\sqrt{\alpha_{1}^{2}+\alpha_{2}^{2}}}|R^{\downarrow}\rangle_{1}|\downarrow\rangle_{1}|\uparrow\rangle_{2}|\uparrow\rangle_{3}\nonumber\\
&+&\frac{\alpha_{1}\alpha_{2}}{\sqrt{\alpha_{1}^{2}+\alpha_{2}^{2}}}|L^{\downarrow}\rangle_{1}|\downarrow\rangle_{1}|\uparrow\rangle_{2}|\uparrow\rangle_{3}\nonumber\\
&+&\frac{\alpha_{1}\alpha_{2}}{\sqrt{\alpha_{1}^{2}+\alpha_{2}^{2}}}|R^{\downarrow}\rangle_{1}|\uparrow\rangle_{1}|\downarrow\rangle_{2}|\uparrow\rangle_{3}\nonumber\\
&+&\frac{\alpha_{2}^{2}}{\sqrt{\alpha_{1}^{2}+\alpha_{2}^{2}}}|L^{\downarrow}\rangle_{1}|\uparrow\rangle_{1}|\downarrow\rangle_{2}|\uparrow\rangle_{3}\nonumber\\
&+&\frac{\alpha_{1}\alpha_{3}}{\sqrt{\alpha_{1}^{2}+\alpha_{2}^{2}}}|R^{\downarrow}\rangle_{1}|\uparrow\rangle_{1}|\uparrow\rangle_{2}|\downarrow\rangle_{3}\nonumber\\
&+&\frac{\alpha_{2}\alpha_{3}}{\sqrt{\alpha_{1}^{2}+\alpha_{2}^{2}}}|L^{\downarrow}\rangle_{1}|\uparrow\rangle_{1}|\uparrow\rangle_{2}|\downarrow\rangle_{3}\nonumber\\
&\rightarrow&\frac{\alpha_{1}^{2}}{\sqrt{\alpha_{1}^{2}+\alpha_{2}^{2}}}|L^{\uparrow}\rangle_{1}|\downarrow\rangle_{1}|\uparrow\rangle_{2}|\uparrow\rangle_{3}\nonumber\\
&-&\frac{\alpha_{1}\alpha_{2}}{\sqrt{\alpha_{1}^{2}+\alpha_{2}^{2}}}|L^{\downarrow}\rangle_{1}|\downarrow\rangle_{1}|\uparrow\rangle_{2}|\uparrow\rangle_{3}\nonumber\\
&-&\frac{\alpha_{1}\alpha_{2}}{\sqrt{\alpha_{1}^{2}+\alpha_{2}^{2}}}|R^{\downarrow}\rangle_{1}|\uparrow\rangle_{1}|\downarrow\rangle_{2}|\uparrow\rangle_{3}\nonumber\\
&+&\frac{\alpha_{2}^{2}}{\sqrt{\alpha_{1}^{2}+\alpha_{2}^{2}}}|R^{\uparrow}\rangle_{1}|\uparrow\rangle_{1}|\downarrow\rangle_{2}|\uparrow\rangle_{3}\nonumber\\
&-&\frac{\alpha_{1}\alpha_{3}}{\sqrt{\alpha_{1}^{2}+\alpha_{2}^{2}}}|R^{\downarrow}\rangle_{1}|\uparrow\rangle_{1}|\uparrow\rangle_{2}|\downarrow\rangle_{3}\nonumber\\
&+&\frac{\alpha_{2}\alpha_{3}}{\sqrt{\alpha_{1}^{2}+\alpha_{2}^{2}}}|R^{\uparrow}\rangle_{1}|\uparrow\rangle_{1}|\uparrow\rangle_{2}|\downarrow\rangle_{3}.\label{evolve1}
\end{eqnarray}
Interestingly, from Eq. (\ref{evolve1}), if the photon is transmitted and in the
output2, the original state collapses to
\begin{eqnarray}
|\Psi'\rangle&=&\frac{\alpha_{1}\alpha_{2}}{\sqrt{\alpha_{1}^{2}+\alpha_{2}^{2}}}|L^{\downarrow}\rangle_{1}|\downarrow\rangle_{1}|\uparrow\rangle_{2}|\uparrow\rangle_{3}\nonumber\\
&+&\frac{\alpha_{1}\alpha_{2}}{\sqrt{\alpha_{1}^{2}+\alpha_{2}^{2}}}|R^{\downarrow}\rangle_{1}|\uparrow\rangle_{1}|\downarrow\rangle_{2}|\uparrow\rangle_{3}\nonumber\\
&+&\frac{\alpha_{1}\alpha_{3}}{\sqrt{\alpha_{1}^{2}+\alpha_{2}^{2}}}|R^{\downarrow}\rangle_{1}|\uparrow\rangle_{1}|\uparrow\rangle_{2}|\downarrow\rangle_{3}.\label{collpase1}
\end{eqnarray}
Then Alice lets her photon pass through the HWP$_{45}$ and PBS$_{2}$. The
HWP$_{45}$   makes
\begin{eqnarray}
|R\rangle\rightarrow\frac{1}{\sqrt{2}}(|H\rangle+|V\rangle),\nonumber\\
|L\rangle\rightarrow\frac{1}{\sqrt{2}}(|H\rangle-|V\rangle),
\end{eqnarray}
and the PBSs  transmit the $|H\rangle$ polarization photon and
reflect $|V\rangle$ polarization photon.

\begin{figure}[!h]
\begin{center}
\includegraphics[width=8cm,angle=0]{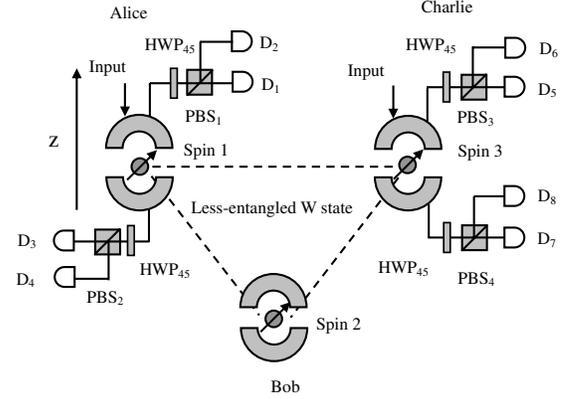}
\caption{A schematic drawing of the basic element of our ECP. The
quantum-dot spin is coupled in optical microcavities.  Input
represents the input port of a photon. Output1 and Oupput2 are the
output ports of the photon after coupled with the electron-spin
system.}
\end{center}
\end{figure}
Finally, if the single-photon detector D$_{3}$ fires, they will get
\begin{eqnarray}
|\Phi_{1}\rangle_{123}&=&\frac{\alpha_{1}\alpha_{2}}{\sqrt{\alpha_{1}^{2}+\alpha_{2}^{2}}}|\downarrow\rangle_{1}|\uparrow\rangle_{2}|\uparrow\rangle_{3}\nonumber\\
&+&\frac{\alpha_{1}\alpha_{2}}{\sqrt{\alpha_{1}^{2}+\alpha_{2}^{2}}}|\uparrow\rangle_{1}|\downarrow\rangle_{2}|\uparrow\rangle_{3}\nonumber\\
&+&\frac{\alpha_{1}\alpha_{3}}{\sqrt{\alpha_{1}^{2}+\alpha_{2}^{2}}}|\uparrow\rangle_{1}|\uparrow\rangle_{2}|\downarrow\rangle_{3}.\label{concentrate1}
\end{eqnarray}
It can be rewritten as
\begin{eqnarray}
 |\Phi_{1}\rangle_{123}&=&\frac{\alpha_{2}}{\sqrt{\alpha_{3}^{2}+2\alpha_{2}^{2}}}|\downarrow\rangle_{1}|\uparrow\rangle_{2}|\uparrow\rangle_{3}\nonumber\\
&+&\frac{\alpha_{2}}{\sqrt{\alpha_{3}^{2}+2\alpha_{2}^{2}}}|\uparrow\rangle_{1}|\downarrow\rangle_{2}|\uparrow\rangle_{3}\nonumber\\
&+&\frac{\alpha_{3}}{\sqrt{\alpha_{3}^{2}+2\alpha_{2}^{2}}}|\uparrow\rangle_{1}|\uparrow\rangle_{2}|\downarrow\rangle_{3}.
\end{eqnarray}

Finally, if the single-photon detector D$_{4}$ fires, they will get
\begin{eqnarray}
 |\Phi_{2}\rangle_{123}&=&-\frac{\alpha_{2}}{\sqrt{\alpha_{3}^{2}+2\alpha_{2}^{2}}}|\downarrow\rangle_{1}|\uparrow\rangle_{2}|\uparrow\rangle_{3}\nonumber\\
&+&\frac{\alpha_{2}}{\sqrt{\alpha_{3}^{2}+2\alpha_{2}^{2}}}|\uparrow\rangle_{1}|\downarrow\rangle_{2}|\uparrow\rangle_{3}\nonumber\\
&+&\frac{\alpha_{3}}{\sqrt{\alpha_{3}^{2}+2\alpha_{2}^{2}}}|\uparrow\rangle_{1}|\uparrow\rangle_{2}|\downarrow\rangle_{3}.\label{concentrate2}
\end{eqnarray}
In order to get $|\Phi_{1}\rangle_{123}$, one of   the three parties, says
Alice, Bob or Charlie should perform a local operation of phase
rotation on her or his  spin.
The success probability is
\begin{eqnarray}
P^{1}_{1}=\frac{|\alpha_{1}|^{2}(|\alpha_{3}|^{2}+2|\alpha_{2}|^{2})}{|\alpha_{1}|^{2}+|\alpha_{2}|^{2}}.\label{probability1}
\end{eqnarray}

On the other hand, after passing through the microcavity, if the
photon is reflected and in the output1, then the Eq. (\ref{evolve1}) collapses to
\begin{eqnarray}
|\Psi''\rangle&=&\frac{\alpha_{1}^{2}}{\sqrt{\alpha_{1}^{2}+\alpha_{2}^{2}}}|L^{\uparrow}\rangle_{1}|\downarrow\rangle_{1}|\uparrow\rangle_{2}|\uparrow\rangle_{3}\nonumber\\
&+&\frac{\alpha_{2}^{2}}{\sqrt{\alpha_{1}^{2}+\alpha_{2}^{2}}}|R,^{\uparrow}\rangle_{1}|\uparrow\rangle_{1}|\downarrow\rangle_{2}|\uparrow\rangle_{3}\nonumber\\
&+&\frac{\alpha_{2}\alpha_{3}}{\sqrt{\alpha_{1}^{2}+\alpha_{2}^{2}}}|R^{\uparrow}\rangle_{1}|\uparrow\rangle_{1}|\uparrow\rangle_{2}|\downarrow\rangle_{3}.\label{collpase2}
\end{eqnarray}
Following the same principle described above, if the D$_{1}$ fires,
they will obtain
\begin{eqnarray}
 |\Phi_{3}\rangle_{123}&=&\frac{\alpha_{1}^{2}}{\sqrt{\alpha_{1}^{4}+\alpha_{2}^{4}+\alpha_{2}^{2}\alpha_{3}^{2}}}|\downarrow\rangle_{1}|\uparrow\rangle_{2}|\uparrow\rangle_{3}\nonumber\\
 &+&\frac{\alpha_{2}^{2}}{\sqrt{\alpha_{1}^{4}+\alpha_{2}^{4}+\alpha_{2}^{2}\alpha_{3}^{2}}}|\uparrow\rangle_{1}|\downarrow\rangle_{2}|\uparrow\rangle_{3}\nonumber\\
 &+&\frac{\alpha_{2}\alpha_{3}}{\sqrt{\alpha_{1}^{4}+\alpha_{2}^{4}+\alpha_{2}^{2}\alpha_{3}^{2}}}|\uparrow\rangle_{1}|\uparrow\rangle_{2}|\downarrow\rangle_{3}.\label{concentrate3}
\end{eqnarray}
if the D$_{2}$ fires, they will obtain
\begin{eqnarray}
 |\Phi_{4}\rangle_{123}&=&-\frac{\alpha_{1}^{2}}{\sqrt{\alpha_{1}^{4}+\alpha_{2}^{4}+\alpha_{2}^{2}\alpha_{3}^{2}}}|\downarrow\rangle_{1}|\uparrow\rangle_{2}|\uparrow\rangle_{3}\nonumber\\
 &+&\frac{\alpha_{2}^{2}}{\sqrt{\alpha_{1}^{4}+\alpha_{2}^{4}+\alpha_{2}^{2}\alpha_{3}^{2}}}|\uparrow\rangle_{1}|\downarrow\rangle_{2}|\uparrow\rangle_{3}\nonumber\\
 &+&\frac{\alpha_{2}\alpha_{3}}{\sqrt{\alpha_{1}^{4}+\alpha_{2}^{4}+\alpha_{2}^{2}\alpha_{3}^{2}}}|\uparrow\rangle_{1}|\uparrow\rangle_{2}|\downarrow\rangle_{3}.\label{concentrate4}
\end{eqnarray}
In order to get $|\Phi_{3}\rangle_{123}$, one of   the parties, says
Alice, Bob or Charlie should perform a local operation of phase
rotation on her or his  spin.

It is interesting to compare $ |\Phi_{1}\rangle_{123}$ with $
|\Phi_{3}\rangle_{123}$. $|\Phi_{1}\rangle_{123}$ only has two
different coefficients, $\frac{\alpha_{2}}{\sqrt{\alpha_{3}^{2}+2\alpha_{2}^{2}}}$
and $\frac{\alpha_{3}}{\sqrt{\alpha_{3}^{2}+2\alpha_{2}^{2}}}$, and the initial
coefficient $\alpha_{1}$ disappears.  But $|\Phi_{3}\rangle_{123}$ still
has three different coefficients. We denote
$\alpha_{1}'\equiv\frac{\alpha_{1}^{2}}{\sqrt{\alpha_{1}^{4}+\alpha_{2}^{4}+\alpha_{2}^{2}\alpha_{3}^{2}}}$,
$\alpha_{2}'\equiv\frac{\alpha_{2}^{2}}{\sqrt{\alpha_{1}^{4}+\alpha_{2}^{4}+\alpha_{2}^{2}\alpha_{3}^{2}}}$
and
$\alpha_{3}'\equiv\frac{\alpha_{2}\alpha_{3}}{\sqrt{\alpha_{1}^{4}+\alpha_{2}^{4}+\alpha_{2}^{2}\alpha_{3}^{2}}}$.
So if Alice obtains  $ |\Phi_{1}\rangle_{123}$, it is successful.
Then she asks Charlie to continue this ECP. Otherwise, she has to
repeat this ECP in a second round. That is, she prepares another single photon of the
form
\begin{eqnarray}
|\Phi\rangle'_{P1}&=&\frac{\alpha_{1}'}{\sqrt{\alpha_{1}'^{2}+\alpha_{2}'^{2}}}|R\rangle_{1}
+\frac{\alpha_{2}'}{\sqrt{\alpha_{1}'^{2}+\alpha_{2}'^{2}}}|L\rangle_{1}.\label{auxiliary2}
\end{eqnarray}
Then she lets this single photon entrance into the microcavity and
couple with the spin. After the photon passing through the microcavity,
following the same principle, if
this single photon is in the output2 and detected by D$_{3}$ or
D$_{4}$, the concentration is successful. Otherwise, if it is in the
output1 and detected by D$_{1}$ or D$_{2}$, the concentration is a
failure. Alice should prepare a third single photon and restart to
perform this ECP until it is successful. So the success probability in the second round is
\begin{eqnarray}
P^{2}_{1}=\frac{|\alpha_{1}|^{4}(|\alpha_{2}|^{2}|\alpha_{3}|^{2}+2|\alpha_{2}|^{4})}{(|\alpha_{1}|^{4}+|\alpha_{2}|^{4})(|\alpha_{1}|^{2}+|\alpha_{2}|^{2})},
\end{eqnarray}
the success probability in the third round is
\begin{eqnarray}
P^{3}_{1}=\frac{|\alpha_{1}|^{8}(|\alpha_{2}|^{6}|\alpha_{3}|^{2}+2|\alpha_{2}|^{8})}{(|\alpha_{1}|^{2}+|\alpha_{2}|^{2})(|\alpha_{1}|^{4}+|\alpha_{2}|^{4})(|\alpha_{1}|^{8}+|\alpha_{2}|^{8})},
\end{eqnarray}
If it is repeated for $K$ times, the success probability is
\begin{eqnarray}
P^{K}_{1}=\frac{|\alpha_{1}|^{2^{K}}(|\alpha_{2}|^{2^{K}-2}|\alpha_{3}|^{2}+2|\alpha_{2}|^{2^{K}})}{(|\alpha_{1}|^{2}+|\alpha_{2}|^{2})(|\alpha_{1}|^{4}+|\alpha_{2}|^{4})\cdots(|\alpha_{1}|^{2^{K}}+|\alpha_{2}|^{2^{K}})}.\nonumber\\
\end{eqnarray}
The total success probability for Alice is
\begin{eqnarray}
P_{1}=P^{1}_{1}+P^{2}_{1}+\cdots+=\sum^{\infty}_{K=1}P^{K}_{1}.
\end{eqnarray}
If Alice is successful, then Charlie start to perform this ECP. His
concentration step is analogy with Alice. In detail, he first
prepares a single photon of the form
\begin{eqnarray}
|\Phi\rangle_{P3}&=&\frac{\alpha_{2}}{\sqrt{\alpha_{3}^{2}+\alpha_{2}^{2}}}|R\rangle_{3}+\frac{\alpha_{3}}{\sqrt{\alpha_{3}^{2}+\alpha_{2}^{2}}}|L\rangle_{3}.\label{auxiliary3}
\end{eqnarray}
Charlie lets his single photon entrance the microcavity and couple
with the spin. Then the state $|\Phi_{1}\rangle_{123}$ combined with
$|\Phi\rangle_{P3}$ evolves as
\begin{eqnarray}
|\Psi\rangle_{1}&=&|\Phi_{1}\rangle_{123}|\Phi\rangle_{P2}=(\frac{\alpha_{2}}{\sqrt{\alpha_{3}^{2}+2\alpha_{2}^{2}}}|\downarrow\rangle_{1}|\uparrow\rangle_{2}|\uparrow\rangle_{3}\nonumber\\
&+&\frac{\alpha_{2}}{\sqrt{\alpha_{3}^{2}+2\alpha_{2}^{2}}}|\uparrow\rangle_{1}|\downarrow\rangle_{2}|\uparrow\rangle_{3}\nonumber\\
&+&\frac{\alpha_{3}}{\sqrt{\alpha_{3}^{2}+2\alpha_{2}^{2}}}|\uparrow\rangle_{1}|\uparrow\rangle_{2}|\downarrow\rangle_{3})\nonumber\\
&(&\frac{\alpha_{2}}{\sqrt{\alpha_{3}^{2}+\alpha_{2}^{2}}}|R^{\downarrow}\rangle_{3}+\frac{\alpha_{3}}{\sqrt{\alpha_{3}^{2}+\alpha_{2}^{2}}}|L^{\downarrow}\rangle_{3})\nonumber\\
&=&\frac{\alpha_{2}\alpha_{3}}{\sqrt{\alpha_{3}^{2}+2\alpha_{2}^{2}}\sqrt{\alpha_{3}^{2}+\alpha_{2}^{2}}}|L^{\downarrow}\rangle_{3}|\downarrow\rangle_{1}|\uparrow\rangle_{2}|\uparrow\rangle_{3}\nonumber\\
&+&\frac{\alpha_{2}^{2}}{\sqrt{\alpha_{3}^{2}+2\alpha_{2}^{2}}\sqrt{\alpha_{3}^{2}+\alpha_{2}^{2}}}|R^{\downarrow}\rangle_{3}|\downarrow\rangle_{1}|\uparrow\rangle_{2}|\uparrow\rangle_{3}\nonumber\\
&+&\frac{\alpha_{2}\alpha_{3}}{\sqrt{\alpha_{3}^{2}+2\alpha_{2}^{2}}\sqrt{\alpha_{3}^{2}+\alpha_{2}^{2}}}|L^{\downarrow}\rangle_{3}|\uparrow\rangle_{1}|\downarrow\rangle_{2}|\uparrow\rangle_{3}\nonumber\\
&+&\frac{\alpha_{2}^{2}}{\sqrt{\alpha_{3}^{2}+2\alpha_{2}^{2}}\sqrt{\alpha_{3}^{2}+\alpha_{2}^{2}}}|R^{\downarrow}\rangle_{3}|\uparrow\rangle_{1}|\downarrow\rangle_{2}|\uparrow\rangle_{3}\nonumber\\
&+&\frac{\alpha_{2}\alpha_{3}}{\sqrt{\alpha_{3}^{2}+2\alpha_{2}^{2}}\sqrt{\alpha_{3}^{2}+\alpha_{2}^{2}}}|R^{\downarrow}\rangle_{3}|\uparrow\rangle_{1}|\uparrow\rangle_{2}|\downarrow\rangle_{3}\nonumber\\
&+&\frac{\alpha_{3}^{2}}{\sqrt{\alpha_{3}^{2}+2\alpha_{2}^{2}}\sqrt{\alpha_{3}^{2}+\alpha_{2}^{2}}}|L^{\downarrow}\rangle_{3}|\uparrow\rangle_{1}|\uparrow\rangle_{2}|\downarrow\rangle_{3}\nonumber\\
&\rightarrow&\frac{\alpha_{2}\alpha_{3}}{\sqrt{\alpha_{3}^{2}+2\alpha_{2}^{2}}\sqrt{\alpha_{3}^{2}+\alpha_{2}^{2}}}|R^{\uparrow}\rangle_{3}|\downarrow\rangle_{1}|\uparrow\rangle_{2}|\uparrow\rangle_{3}\nonumber\\
&-&\frac{\alpha_{2}^{2}}{\sqrt{\alpha_{3}^{2}+2\alpha_{2}^{2}}\sqrt{\alpha_{3}^{2}+\alpha_{2}^{2}}}|R^{\downarrow}\rangle_{3}|\downarrow\rangle_{1}|\uparrow\rangle_{2}|\uparrow\rangle_{3}\nonumber\\
&+&\frac{\alpha_{2}\alpha_{3}}{\sqrt{\alpha_{3}^{2}+2\alpha_{2}^{2}}\sqrt{\alpha_{3}^{2}+\alpha_{2}^{2}}}|R^{\uparrow}\rangle_{3}|\uparrow\rangle_{1}|\downarrow\rangle_{2}|\uparrow\rangle_{3}\nonumber\\
&-&\frac{\alpha_{2}^{2}}{\sqrt{\alpha_{3}^{2}+2\alpha_{2}^{2}}\sqrt{\alpha_{3}^{2}+\alpha_{2}^{2}}}|R^{\downarrow}\rangle_{3}|\uparrow\rangle_{1}|\downarrow\rangle_{2}|\uparrow\rangle_{3}\nonumber\\
&+&\frac{\alpha_{2}\alpha_{3}}{\sqrt{\alpha_{3}^{2}+2\alpha_{2}^{2}}\sqrt{\alpha_{3}^{2}+\alpha_{2}^{2}}}|L^{\uparrow}\rangle_{3}|\uparrow\rangle_{1}|\uparrow\rangle_{2}|\downarrow\rangle_{3}\nonumber\\
&-&\frac{\alpha_{3}^{2}}{\sqrt{\alpha_{3}^{2}+2\alpha_{2}^{2}}\sqrt{\alpha_{3}^{2}+\alpha_{2}^{2}}}|L^{\downarrow}\rangle_{3}|\uparrow\rangle_{1}|\uparrow\rangle_{2}|\downarrow\rangle_{3}.\nonumber\\\label{evolve2}
\end{eqnarray}
From Eq. (\ref{evolve2}), after the photon  passing through the
microcavity, if the photon is in the output1, then above state
collapses to
\begin{eqnarray}
|\Psi\rangle'_{1}&=&\frac{\alpha_{2}\alpha_{3}}{\sqrt{\alpha_{3}^{2}+2\alpha_{2}^{2}}\sqrt{\alpha_{3}^{2}+\alpha_{2}^{2}}}|R^{\uparrow}\rangle_{3}|\downarrow\rangle_{1}|\uparrow\rangle_{2}|\uparrow\rangle_{3}\nonumber\\
&+&\frac{\alpha_{2}\alpha_{3}}{\sqrt{\alpha_{3}^{2}+2\alpha_{2}^{2}}\sqrt{\alpha_{3}^{2}+\alpha_{2}^{2}}}|R^{\uparrow}\rangle_{3}|\uparrow\rangle_{1}|\downarrow\rangle_{2}|\uparrow\rangle_{3}\nonumber\\
&+&\frac{\alpha_{2}\alpha_{3}}{\sqrt{\alpha_{3}^{2}+2\alpha_{2}^{2}}\sqrt{\alpha_{3}^{2}+\alpha_{2}^{2}}}|L^{\uparrow}\rangle_{3}|\uparrow\rangle_{1}|\uparrow\rangle_{2}|\downarrow\rangle_{3}.\nonumber\\
\end{eqnarray}
It can be rewritten as
\begin{eqnarray}
|\Psi\rangle'_{1}&=&\frac{1}{\sqrt{3}}(|R^{\uparrow}\rangle_{3}|\downarrow\rangle_{1}|\uparrow\rangle_{2}|\uparrow\rangle_{3}\nonumber\\
&+&|R^{\uparrow}\rangle_{3}|\uparrow\rangle_{1}|\downarrow\rangle_{2}|\uparrow\rangle_{3}+|L^{\uparrow}\rangle_{3}|\uparrow\rangle_{1}|\uparrow\rangle_{2}|\downarrow\rangle_{3}).\nonumber\\
\end{eqnarray}
Finally, after the photon passing through the HWP$_{45}$ and
PBS$_{2}$, if D$_{5}$ fires, the remained state is essentially the
maximally entangled W state
\begin{eqnarray}
 |\Phi_{5}\rangle_{123}&=&\frac{1}{\sqrt{3}}(|\downarrow\rangle_{1}|\uparrow\rangle_{2}|\uparrow\rangle_{3}\nonumber\\
&+&|\uparrow\rangle_{1}|\downarrow\rangle_{2}|\uparrow\rangle_{3}+|\uparrow\rangle_{1}|\uparrow\rangle_{2}|\downarrow\rangle_{3}).
\end{eqnarray}
If  D$_{6}$ fires, the remained state is
\begin{eqnarray}
 |\Phi_{6}\rangle_{123}&=&\frac{1}{\sqrt{3}}(|\downarrow\rangle_{1}|\uparrow\rangle_{2}|\uparrow\rangle_{3}\nonumber\\
&+&|\uparrow\rangle_{1}|\downarrow\rangle_{2}|\uparrow\rangle_{3}-|\uparrow\rangle_{1}|\uparrow\rangle_{2}|\downarrow\rangle_{3}).
\end{eqnarray}
They can obtain $|\Phi_{5}\rangle_{123}$ by performing a local operation of the phase rotation
on one of the spin.

On the other hand, if the photon is in the output2 and leads the D$_{7}$
fire, the Eq. (\ref{evolve2}) collapses to
\begin{eqnarray}
 |\Phi_{7}\rangle_{123}&=&\frac{\alpha_{2}^{2}}{\sqrt{\alpha_{3}^{2}+2\alpha_{2}^{2}}\sqrt{\alpha_{3}^{2}+\alpha_{2}^{2}}}|\downarrow\rangle_{1}|\uparrow\rangle_{2}|\uparrow\rangle_{3}\nonumber\\
 &+&\frac{\alpha_{2}^{2}}{\sqrt{\alpha_{3}^{2}+2\alpha_{2}^{2}}\sqrt{\alpha_{3}^{2}+\alpha_{2}^{2}}}|\uparrow\rangle_{1}|\downarrow\rangle_{2}|\uparrow\rangle_{3}\nonumber\\
 &+&\frac{\alpha_{3}^{2}}{\sqrt{\alpha_{3}^{2}+2\alpha_{2}^{2}}\sqrt{\alpha_{3}^{2}+\alpha_{2}^{2}}}|\uparrow\rangle_{1}|\uparrow\rangle_{2}|\downarrow\rangle_{3}.\label{collpase3}
\end{eqnarray}
It can be written as
\begin{eqnarray}
 |\Phi_{7}\rangle_{123}&=&\frac{\alpha_{2}^{2}}{\sqrt{2\alpha_{2}^{4}+\alpha_{3}^{4}}}|\downarrow\rangle_{1}|\uparrow\rangle_{2}|\uparrow\rangle_{3}\nonumber\\
 &+&\frac{\alpha_{2}^{2}}{\sqrt{2\alpha_{2}^{4}+\alpha_{3}^{4}}}|\uparrow\rangle_{1}|\downarrow\rangle_{2}|\uparrow\rangle_{3}\nonumber\\
 &+&\frac{\alpha_{3}^{2}}{\sqrt{2\alpha_{2}^{4}+\alpha_{3}^{4}}}|\uparrow\rangle_{1}|\uparrow\rangle_{2}|\downarrow\rangle_{3}.\label{collpase4}
\end{eqnarray}

Otherwise, if D$_{8}$ fires, the Eq. (\ref{evolve2}) collapses to
\begin{eqnarray}
 |\Phi_{8}\rangle_{123}&=&\frac{\alpha_{2}^{2}}{\sqrt{2\alpha_{2}^{4}+\alpha_{3}^{4}}}|\downarrow\rangle_{1}|\uparrow\rangle_{2}|\uparrow\rangle_{3}\nonumber\\
 &+&\frac{\alpha_{2}^{2}}{\sqrt{2\alpha_{2}^{4}+\alpha_{3}^{4}}}|\uparrow\rangle_{1}|\downarrow\rangle_{2}|\uparrow\rangle_{3}\nonumber\\
 &-&\frac{\alpha_{3}^{2}}{\sqrt{2\alpha_{2}^{4}+\alpha_{3}^{4}}}|\uparrow\rangle_{1}|\uparrow\rangle_{2}|\downarrow\rangle_{3}.\label{collpase5}
\end{eqnarray}

They can also obtain $|\Phi_{7}\rangle_{123}$, by performing a local operation of phase
rotation on one of the  spin. Eqs. (\ref{collpase4}) and
(\ref{collpase5}) are both lesser-entangled W states, which have the
same form of $|\Phi_{1}\rangle_{123}$. That is, if Charlie obtains
Eq. (\ref{collpase4}), he can repeat this ECP and obtain the
maximally entangled W state in a second round.
They can also obtain the success probability for Charlie as
\begin{eqnarray}
P^{1}_{2}&=&\frac{3|\alpha_{2}|^{2}|\alpha_{3}|^{2}}{(|\alpha_{3}|^{2}+|\alpha_{2}|^{2})(|\alpha_{3}|^{2}+2|\alpha_{2}|^{2})},\nonumber\\\label{probability2}
P^{2}_{2}&=&\frac{3|\alpha_{2}|^{4}|\alpha_{3}|^{4}}{(|\alpha_{3}|^{2}+|\alpha_{2}|^{2})(|\alpha_{3}|^{4}+|\alpha_{2}|^{4})(|\alpha_{3}|^{2}+2|\alpha_{2}|^{2})}\nonumber\\
&\cdots&\nonumber\\
P^{K}_{2}&=&\frac{3|\alpha_{2}|^{2^{K}}|\alpha_{3}|^{2^{K}}}{(|\alpha_{3}|^{2}+|\alpha_{2}|^{2})(|\alpha_{3}|^{4}+|\alpha_{2}|^{4})\cdots(|\alpha_{3}|^{2^{K}}+|\alpha_{2}|^{2^{K}})}\nonumber\\
&\times&\frac{1}{(|\alpha_{3}|^{2}+2|\alpha_{2}|^{2})}.
\end{eqnarray}
The total success probability for Charlie is
\begin{eqnarray}
P_{2}=P^{1}_{2}+P^{2}_{2}+\cdots+=\sum^{\infty}_{K=1}P^{K}_{2}.
\end{eqnarray}

\section{Success probability and experiment feasibilities}
Thus far, we have briefly explained this ECP. We can calculate the
success probability for both Alice and Charlie. 
Suppose that Alice and Charlie repeat this ECP for $K$ times, the total success probability is
\begin{eqnarray}
P_{t}=\sum^{\infty}_{K=1}P^{K}_{1}\sum^{\infty}_{K=1}P^{K}_{2}.
\end{eqnarray}
 If both Alice and Charlie perform this protocol only one time, 
the total success probability
\begin{eqnarray}
P^{1}_{t}=P^{1}_{1}P^{1}_{2}=\frac{3|\alpha_{1}|^{2}|\alpha_{2}|^{2}|\alpha_{3}|^{2}}{(|\alpha_{1}|^{2}+|\alpha_{2}|^{2})(|\alpha_{3}|^{2}+|\alpha_{2}|^{2})}.\label{probability3}
\end{eqnarray}

\begin{figure}[!h]
\begin{center}
\includegraphics[width=8cm,angle=0]{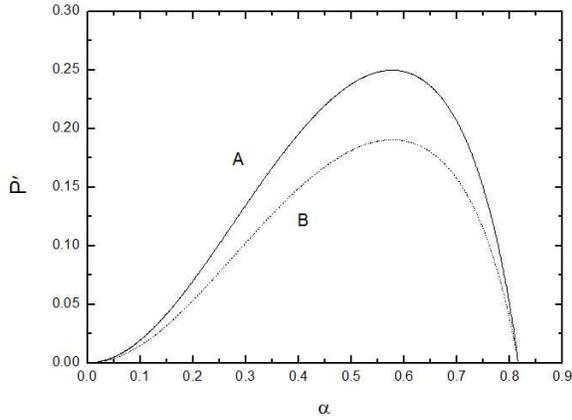}
\caption{Success probability $P'$ of obtaining a maximally entangled
W state after performing this ECP is altered with the initial
coefficient $\alpha_{1}(0,\sqrt{\frac{2}{3}})$. We chose
$\alpha_{2}=\frac{1}{\sqrt{3}}$. Curve A is the idea case with  no
leakage. Curve B is the success probability with
 $\kappa_{s}=0.1\kappa$, $g=0.5\kappa$
and $\gamma=0.1\kappa$.}
\end{center}
\end{figure}

 Actually, the realization of this ECP relies on the efficiency of hybrid parity
check for electrons and photon described in Sec. II. By solving the
Heisenberg equations of motion for the cavity-field operator and the
trion dipole operator in weak excitation approximation, we can
calculate the practical transmission and reflection coefficients.
Similar to Ref. \cite{hu2}, we denote $\omega_{0}$, $\omega_{c}$ and
$\omega_{X^{-}}$ as the frequencies of the input photon, cavity
mode, and the spin-dependent optical transition, respectively. In
the approximation of weak excitation, the reflection and
transmission coefficients can be written as
\begin{eqnarray}
r(w)&=&1+t(\omega),\nonumber\\
t(\omega)&=&\frac{-\kappa[i(\omega_{X^{-}}-\omega)+\frac{\gamma}{2}]}{[i(\omega_{X^{-}}-\omega)+\frac{\gamma}{2}][i(\omega_{c}-\omega)+\kappa+\frac{\kappa_{s}}{2}+g^{2}]},\nonumber\\
\end{eqnarray}
where $g$ represents the coupling constant. $\frac{\gamma}{2}$ is the
$X^{-} $ dipole decay rate. $\kappa$ and $\kappa_{s}/2$ are the
cavity field decay rate into the input and output modes and the leaky
rate, respectively.
\begin{figure}[!h]
\begin{center}
\includegraphics[width=8cm,angle=0]{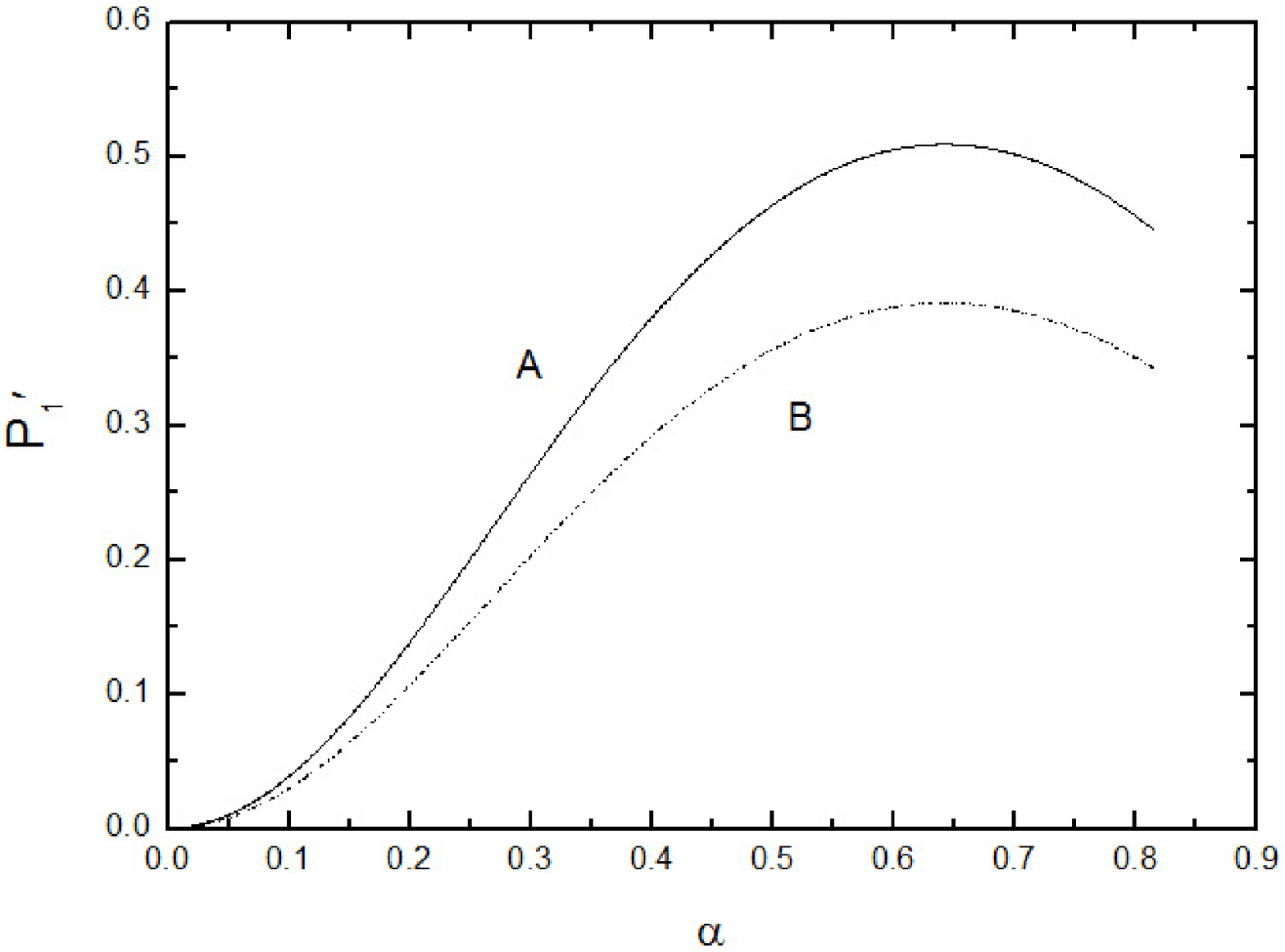}
\caption{Success probability $P'_{1}$ for Alice performing this ECP
shown in Eq. (\ref{probability4}). We chose
$\alpha_{2}=\frac{1}{\sqrt{3}}$ and  $\alpha_{1}\in(0,\sqrt{\frac{2}{3}})$.
Curve A is the idea case with no leakage and Curve B represents the
success probability with $\kappa_{s}=0.5\kappa$, $g=0.5\kappa$, and
$\gamma=0.1\kappa$.}
\end{center}
\end{figure}
If we consider the condition that the resonant interaction with
$\omega_{c}=\omega_{X^{-}}=\omega_{0}$, and $g=0$, we can get the
reflection and transmission coefficients as
\begin{eqnarray}
r_{0}(\omega)=\frac{i(\omega_{0}-\omega)+\frac{\kappa_{s}}{2}}{i(\omega_{0}-\omega)+\frac{\kappa_{s}}{2}+\kappa},\nonumber\\
t_{0}(\omega)=\frac{-\kappa}{i(\omega_{0}-\omega)+\frac{\kappa_{s}}{2}+\kappa}.
\end{eqnarray}
The transmission and reflection operators can be rewritten as
\begin{eqnarray}
\hat{t}(\omega)=t_{0}(\omega)(|R\rangle\langle R|\otimes
|\uparrow\rangle\langle\uparrow|+|L\rangle\langle L|\otimes
|\downarrow\rangle\langle\downarrow|)\nonumber\\
+t(\omega)(|R\rangle\langle R|\otimes
|\uparrow\rangle\langle\uparrow|+|L\rangle\langle L|\otimes
|\downarrow\rangle\langle\downarrow|),\nonumber\\
\hat{r}(\omega)=r_{0}(\omega)(|R\rangle\langle R|\otimes
|\uparrow\rangle\langle\uparrow|+|L\rangle\langle L|\otimes
|\downarrow\rangle\langle\downarrow|)\nonumber\\
+r(\omega)(|R\rangle\langle R|\otimes
|\uparrow\rangle\langle\uparrow|+|L\rangle\langle L|\otimes
|\downarrow\rangle\langle\downarrow|).
\end{eqnarray}

\begin{figure}[!h]
\begin{center}
\includegraphics[width=8cm,angle=0]{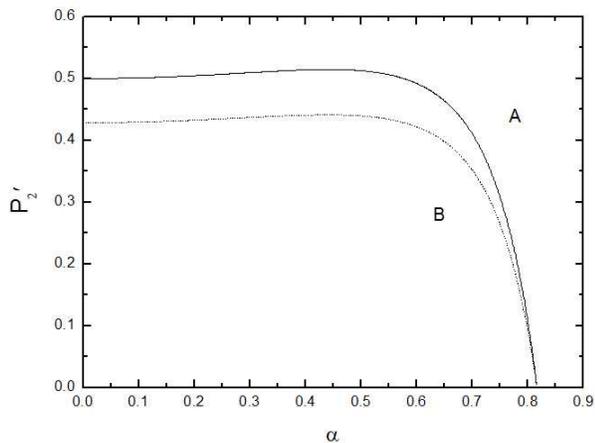}
\caption{Success probability $P'_{2}$ for Charlie performing this
ECP shown in Eq. (\ref{probability5}). We chose
$\alpha_{2}=\frac{1}{\sqrt{3}}$ and $\alpha_{1}\in(0,\sqrt{\frac{2}{3}})$.
Curve A is the idea case with no leakage and Curve B represents the
success probability with $\kappa_{s}=0.5\kappa$, $g=0.5\kappa$, and
$\gamma=0.1\kappa$.}
\end{center}
\end{figure}
In a practical experiment, we let the Alice and Charlie only perform this ECP for one time, so the success probability for Alice can be rewritten as
\begin{eqnarray}
P'_{1}=\frac{|\alpha_{1}|^{2}(|\alpha_{3}|^{2}+2|\alpha_{2}|^{2})}{|\alpha_{1}|^{2}+|\alpha_{2}|^{2}}\frac{|t_{0}(\omega)|}{\sqrt{|t_{0}(\omega)|^{2}+|t(\omega)|^{2}}}.\label{probability4}
\end{eqnarray}
The practical success probability for Charlie can be rewritten as
\begin{eqnarray}
P'_{2}=\frac{3|\alpha_{2}|^{2}|\alpha_{3}|^{2}}{(|\alpha_{3}|^{2}+|\alpha_{2}|^{2})(|\alpha_{3}|^{2}+2|\alpha_{2}|^{2})}\frac{|r(\omega)|}{\sqrt{|r_{0}(\omega)|^{2}+|r(\omega)|^{2}}}.\nonumber\\\label{probability5}
\end{eqnarray}
The total probability $P'$ can be rewritten as
\begin{eqnarray}
P'=P'_{1}P'_{2}=\frac{3|\alpha_{1}|^{2}|\alpha_{2}|^{2}|\alpha_{3}|^{2}}{(|\alpha_{1}|^{2}+|\alpha_{2}|^{2})(|\alpha_{3}|^{2}+|\alpha_{2}|^{2})}\nonumber\\
\frac{|t_{0}(\omega)|r_{0}(\omega)|}{\sqrt{(|t_{0}(\omega)|^{2}+|t(\omega)|^{2})(|r_{0}(\omega)|^{2}+|r(\omega)|^{2}})}.\label{probability6}
\end{eqnarray}

\begin{figure}[!h]
\begin{center}
\includegraphics[width=8cm,angle=0]{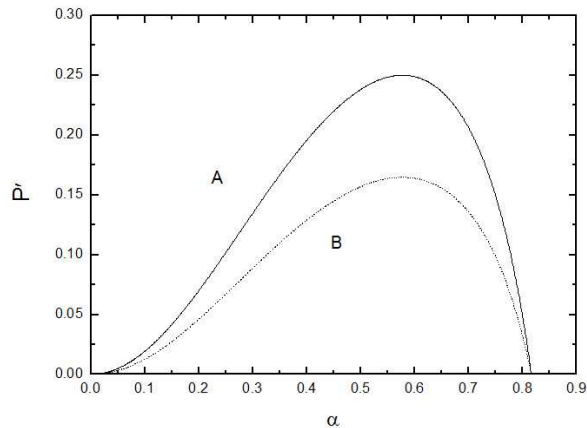}
\caption{Success probability $P'$ of obtaining a maximally entangled
W state after performing this ECP is altered with the initial
coefficient $\alpha_{1}\in(0,\sqrt{\frac{2}{3}})$. We chose
$\alpha_{2}=\frac{1}{\sqrt{3}}$.  Curve A is the idea case with no
leakage. Curve B is the success probability with
 $\kappa_{s}=0.5\kappa$, $g=0.5\kappa$
and $\gamma=0.1\kappa$.}
\end{center}
\end{figure}
We calculated the success probability of this ECP for different
$\alpha_{1}$ in the coupled system. Fig. 3 illustrates the success
probability $P'$ shown in Eq. (\ref{probability6}) for obtaining the
maximally entangled W state. We chose $\alpha_{2}=\frac{1}{\sqrt{3}}$ and
changed  $\alpha_{1}\in(0,\sqrt{\frac{2}{3}})$. It is shown that in the
idea case without leakage, the success probability can reach 0.25
when $\alpha_{1}=\alpha_{2}=\alpha_{3}=\frac{1}{\sqrt{3}}$. But in a practical
condition that $\kappa_{s}=0.1\kappa$, the max value
$P'\approx0.18$, when  $\alpha_{1}=\alpha_{2}=\alpha_{3}=\frac{1}{\sqrt{3}}$. We
also calculated the success probability when $\kappa_{s}=0.5\kappa$,
shown in Fig. 4, Fig. 5 and Fig. 6. In Fig. 4, it is the success
probability for Alice performing this ECP one time in both the idea
case and with the leakage $\kappa_{s}=0.5\kappa$. It is shown that
in the idea case, the max value can reach 0.5 and it can reach
$P'_{1}\approx0.4$ when $\kappa_{s}=0.5\kappa$. Interestingly, the
success probability shown in Eqs. (\ref{probability2}) and
(\ref{probability5})  for Charlie remains almost unchanged when
$\alpha_{1}\in(0,0.6)$ with $P_{2}\approx0.5$ and $P'_{2}\approx0.43$.
But both $P_{2}$ and $P'_{2}$  decrease rapidly when
$\alpha_{1}\in(0.6,\sqrt{\frac{2}{3}})$. Actually, when
$\alpha_{2}=\frac{1}{\sqrt{3}}$, Eqs. (\ref{probability2}) and
(\ref{probability5}) can be simplified as
\begin{eqnarray}
P_{2}=\frac{\frac{2}{3}-|\alpha_{1}|^{2}}{(1-|\alpha_{1}|^{2})(\frac{2}{3}-|\alpha_{1}|^{2})}\label{probability7},
\end{eqnarray}
and
\begin{eqnarray}
P'_{2}=\frac{\frac{2}{3}-|\alpha_{1}|^{2}}{(1-|\alpha_{1}|^{2})(\frac{2}{3}-|\alpha_{1}|^{2})}\frac{|r(\omega)|}{\sqrt{|r_{0}(\omega)|^{2}+|r(\omega)|^{2}}}.\label{probability8}
\end{eqnarray}
In Fig. 6 we calculated the success probability for obtaining a
maximally entangled W state when $\kappa_{s}=0.5\kappa$. Compared
with Fig. 3, the success probability decreases. The max value is
about $P'\approx0.16$.

\section{discussion and summary}
In this ECP, the system can be realized in a self-assembled GaAs
quantum dot or an InAs interface quantum dot in micropillar
microcavities. Therefore, the long coherence of quantum dots and the
strong coupling of the quantum dots with the cavity are required. It
is reported that the coupling strength is acceptable for  $g=0.5$ in
a microcavity with a diameter of $d=1.5 \mu m$ with the cavity
leakage\cite{QD1}. In 2008, Press \emph{et al.} demonstrated
 the optical initialization, rotation by arbitrary angle and projective measurement of an electron spin in a quantum dot\cite{QD4}.
 They showed that coherence time is 3.0 $\mu s$ and about $10^{5}$ operations can be achieved within the qubit's coherent time.
 Greilich \emph{et al.} also reported their experiment about ultrafast optical rotation of spins about arbitrary axes
  on a picosecond timescale using laser pulses as control fields\cite{QD5}.
 Current experiment also demonstrated that the spin coherent
time $T^{h}_{2}> 100 ns$  due to the suppressed electron-photon
interaction and the lack of hole-nuclear hyperfine interaction\cite{QD2}.
The cooling and fast coherent control of hole-spin
states were also reported\cite{QD2,QD3}. In 2010, Press \emph{et al.} increased decoherence time of a single quantum dot
  electron spin from nonaseconds to microseconds using  ultrafast all-optical spin echo technique\cite{QD6}.

In summary, we exploit the single photons to concentrate the
less-entangled W state for the charge qubits confined in the quantum
dots. This ECP is quite different from the other protocols because
this ECP is performed between different physical qubits, i.e. the
charge qubits and the photons, while other protocols unusually use
the same physical qubits. It provides us a good way to realize such
ECP. Moreover, during this ECP, we only require one pair of
less-entangled W states while the conventional ECPs should resorts
two same copies of such states. Therefore, this ECP seems more
optimal. On the other hand, this protocol can be repeated to obtain
a high success probability by consuming some single photons. Our
protocol may be useful and  flexible in current quantum information
processing.

\section*{ACKNOWLEDGEMENTS}
This work was supported by the National Natural Science Foundation
of China under Grant No. 11104159,  Scientific Research Foundation
of Nanjing University of Posts and Telecommunications under Grant
No. NY211008, University Natural Science Research Foundation of
JiangSu Province under Grant No. 11KJA510002,  the open research
fund of Key Lab of Broadband Wireless Communication and Sensor
Network Technology (Nanjing University of Posts and
Telecommunications), Ministry of Education, China, and A Project
Funded by the Priority Academic Program Development of Jiangsu
Higher Education Institutions.

\end{document}